\documentstyle[epsfig]{mn}
\begin{document}


\title{The IRAS PSCz Dipole}
\author
[M.Rowan-Robinson {\it et al.}]
{M. Rowan-Robinson$^1$, J.Sharpe$^1$, S.J.Oliver$^1$, 
O.Keeble$^1$, A.Canavezes$^1$,
\vspace*{0.2cm}\\{\LARGE  
W.Saunders$^2$, A.N.Taylor$^2$, H.Valentine$^2$, 
C.S.Frenk$^3$,
G.P.Efstathiou$^4$,
}\vspace*{0.2cm}\\{\LARGE 
R.G.McMahon$^4$, S.D.M.White$^5$,W.Sutherland$^6$, 
H.Tadros$^6$, 
S.Maddox$^7$
}\\
$^1$Astrophysics Group, Imperial College London, Blackett Laboratory,
Prince Consort Road, London SW7 2BZ;\\ 
$^2$Institute for Astronomy, Edinburgh;\\
$^3$Dept of Physics, University of Durham;\\ 
$^4$Institute of Astronomy, Cambridge;\\
$^5$Max-Planck-Institut f\"ur Astrophysik, Karl-Schwarzschild-Strasse 1, Garching bei Munchen, 
Germany D-85740;\\
$^6$Dept of Astrophysio
cs, Oxford;\\
$^7$School of Physics and Astronomy, University of Nottingham, Nottingham NG7 2RD. \\
}
\maketitle
\begin{abstract} 

We use the PSCz IRAS galaxy redshift survey to analyze the cosmological galaxy dipole out to a distance
300 h$^{-1}$ Mpc.
The masked area is filled in three different ways, firstly by sampling the whole sky at random,
 secondly by using neighbouring areas to fill a masked region, and thirdly using a spherical
harmonic analysis.  The method of treatment
of the mask is found to have a significant effect on the final calculated dipole.

The conversion from redshift space to real space is accomplished by using an analytical
model of the cluster and void distribution, based on 88 nearby groups, 854 clusters and
163 voids, with some of the clusters and all of the voids found from the PSCz database.

The dipole for the whole PSCz sample appears to have converged within a distance of 200 $h^{-1}$Mpc and
yields a value for  $\beta = \Omega^{0.6}/b$ = 0.75 (+0.11,-0.08),
consistent with earlier determinations from IRAS samples by a variety of methods.
For b = 1, the $2-\sigma$ range for $\Omega_{o}$ is 0.43-1.02.  

The direction of the dipole is within 13$^o$ of the CMB dipole, the main uncertainty in
direction being associated with the masked area behind the Galactic plane.  The improbability
that further major contributions to the dipole amplitude will come from volumes larger
than those surveyed here means that the question of the origin of the CMB dipole is
essentially resolved.

\end{abstract}


\section{Introduction}

IRAS all-sky redshift surveys have been used to study the cosmological dipole by Strauss and
Davis (1988, 1.94 Jy sample), 
Rowan-Robinson et al (1990, QDOT sample) and Strauss et al (1992, 1.2 Jy sample).  In 
this paper we report results from the new IRAS PSCz Redshift Survey (Saunders et al 1995, 1997, 1999 in preparation), 
which includes
redshifts for over 15000 IRAS galaxies with 60 $\mu$m fluxes brighter than 0.6 Jy.

The cosmic microwave background (CMB) dipole is interpreted as a Doppler 
effect arising from the motion of the Local Group of galaxies through the cosmological
frame.  The IRAS studies cited above showed clearly for the first time that this motion
could be explained as due to the net attraction of matter, distributed broadly like
the galaxies, within 150 $h^{-1}$ Mpc.  If the galaxy distribution gives an unbiassed
picture of the total matter distribution, values for the cosmological density parameter
$\Omega_{o}$ of 0.5-1.0 and 0.22-0.76 ($1-\sigma$ range) were found by Rowan-Robinson et 
al (1990) and Strauss et al (1992), respectively.  If the galaxy distribution is linearly 
biased with respect to the matter distribution, so that

($\delta \rho/\rho)|_{gal}$ = b  ($\delta \rho/\rho)|_{tot}$,

then it is preferable to work in terms of the parameter  $\beta = \Omega^{0.6}/b$.  The values for
$\beta$ found from a variety of large-scale structure studies using IRAS galaxy samples have
been summarized by Dekel (1994), Strauss and Willick (1995) and Rowan-Robinson (1997).  In
the latter review it was shown that whether averaged for different methods using the same
IRAS sample, or averaged for different IRAS samples using the same method, the values of 
$\beta$ were all consistent with a mean value of 0.80 $\pm$ 0.15.

One of the main interests of the present study is whether the IRAS dipole does in fact converge
by 150 $h^{-1}$ Mpc, or whether major contributions to the motion of the Local Group arise at
larger distances, as proposed by Raychaudhury (1989), Scaramella et al (1989), and Plionis 
and Valdarnini (1991).  The convergence of cosmological dipoles has been discussed by Vittorio 
and Juskiewicz (1987, Juskiewicz et al (1990), Lahav et al (1990), Peacock (1992) and Strauss
et al (1992) (see section 7 below).

Other results from the PSCz survey have been presented by Canavezes et al (1998), Sutherland
et al (1999), Tadros et al (1999), Branchini et al (1999), Schmoldt et al (1999), and Sharpe 
et al (1999).

\section{The PSCz sample}
The IRAS PSCz sample contains 15459 galaxies brighter than 0.6 Jy at 60 $\mu$m in 84 $\%$ 
of the sky.  Saunders et al (1995, 1997, 1999 in preparation) have described the construction of the PSCz sample,
which is based on the IRAS galaxy catalogue of Rowan-Robinson et al (1990).  Improved
60 $\mu$m fluxes have been derived for extended sources.  In addition to the redshifts 
measured in the QDOT survey (Rowan-Robinson et al 1990, Lawrence et al 1999) and in the
1.2 Jy survey (Strauss et al 1992), and those galaxies with redshifts already determined 
in the literature, we measured the redshifts of a further 4500 galaxies.  The data
reduction for these observations was performed primarily by Keeble (1996).  The redshift
distribution for the sample is shown in Fig 1, together with the selection function
assumed in this paper, which is derived from a luminosity function of the form assumed
by Saunders et al (1989) with $log L_* = 8.45, \sigma = 0.711, \alpha = 1.09, C =0.0308$ 
(all assuming $H_o$ = 100) and luminosity evolution of the form  $L_* \propto exp -3t/t_o$
(the normalisation has been increased by 10$\%$ to give the correct total number of sources).
The identifications are believed to be complete over the unmasked sky to V = 30,000 km/s (but see section 3).
We have used a simple window function which is 1 for 4 $h^{-1} Mpc < d <$ 300 $h^{-1} Mpc$ and zero
otherwise.  Only galaxies with 60 $\mu$m luminosities in the range $log_{10} L_{60} = 8 - 13$ are 
used in the dipole calculation.  Strauss et al (1992) discuss the effects of different assumptions 
about the window function, which are small compared with other factors.

\section{Treatment of mask}

Because parts of the sky were either not covered by IRAS or are too severely confused by
emission from our Galaxy or the Magellanic Clouds to be useful for extragalactic studies,
part of the sky (16 $\%$) is masked.  This basic mask has been defined by the I(100$\mu$m) =
25 MJy$/$sr contour.  There have been a variety of approaches to how to correct for the
masked sky.  Strauss and Davis (1988) and Rowan-Robinson et al (1990) 
filled the masked area with a Poissonian distribution of sources.  Lynden-Bell et al
(1989) suggested that the sky at $|b| < 15^o$ be filled by cloning adjacent latitude strips
(a similar method is employed by Yahil et al 1991, Strauss et al 1992).
Scharf et al (1992) emphasized the power of a spherical harmonic approach in bridging what
is known about the unmasked sky across the masked areas.  Lahav et al (1994) refined the
spherical harmonic approach by using Wiener filtering.  A major part of the present study
has been the exploration of the effects of different approaches to filling the masked areas
on the results.

Our first attempt to treat the mask was as follows.
The mask is defined in terms of IRAS 'lune bins', a binning of the sky in 
1 sq deg areas defined in ecliptic coordinates.  We divided the sky into
413 areas each of approximately 100 sq deg and compiled statistics on the
proportion of each of these 100 sq deg areas which lie in the mask.  Areas
in which more than 0 $\%$ but less than 25 $\%$ of the area is masked are then filled
by resampling the data in that area to select fluxes and velocities, and then
placing the coordinates at random in the masked area.

Where more than 25 $\%$ of the area is masked, or the
number of sources falls below 10 (compared with an average per 100 sq deg area
of 40), the areas are filled using one of two algorithms: (A) the area is filled with
flux-velocity pairs randomly selected from the whole data set, to the average 
density 
over the sky, at random locations within the masked area, (B) the area is filled 
with
flux-velocity pairs randomly selected from two neighbouring bins which are at
least 75 $\%$ full, at random locations within the masked area.
Both methods retain the radial density structure of the PSCz data set, but
method (B) assumes that there is strong correlation over scales of 10 degrees,
whereas method (A) assumes no correlation.  The truth is likely to lie
between these two assumptions.

Figs 2-4 show the PSCz data, where the zones of complete avoidance can be
clearly seen, and the augmented data with the mask filled by these two methods.
Both distributions look reasonably convincing.  Where the masked areas are
filled from neighbouring bins, spatial structure can be seen to bridge the
Galactic plane more dramatically.

However both these treatments were found to result in strong changes to the dipole amplitude
and direction, especially for the x- and y-components (where the z-axis is towards the Galactic 
pole and the x-axis towards the Galactic centre), at distances between 200 and 300 $h^{-1}$ Mpc.
This drift in the dipole components can be seen in the preliminary results from this study
shown by Saunders et al (1998, Fig 3).  These changes did not appear to correspond to 
actual structures in the galaxy distribution.
To test whether this could be due
to anisotropic incompleteness in the PSCz survey at large distances, we show in Fig 5
the sky distribution of PSCz galaxies for which we do not have redshifts.  It does
appear that there is a strong concentration of PSCz sources which we believe to be galaxies,
but for which we do not have redshifts, towards the Galactic anticentre.  These are likely
to be preferentially galaxies at larger distances.  Thus in filling the mask with clones of
galaxies from more complete regions, we are generating a spurious component in the positive
x-direction, seen as the upwards drift of the x-component of the dipole at large distances.

To compensate for this effect we model the incompleteness by supposing that the
completeness limit varies smoothly from 15,000 km$/$s at I(100) = 25 MJy$/$sr to 30,000
km$/$s at I(100) = 12.5 MJy$/$sr.  The sample is then completed to 30,000 km$/$s through this
zone using a new method based on spherical harmonics (Saunders 
et al, 1999, in prep). The galaxies are assumed
to be Poisson-sampled from a lognormal underlying distribution, and the
harmonics giving the maximum likelihood for the galaxy distribution
outside the mask determined. The method allows interpolation on mildly
non-linear scales, where the linear Weiner-filtering method (e.g. Lahav
etal 1994) breaks down. A regularisation term, equivalent to that used in 
Weiner filtering, is added to the likelihood to damp out noise
fluctuations.
 
The masked areas (I(100) $>$ 25 MJy$/$sr) are then filled in two ways:
(C) with a random distribution of IRAS galaxies,
(D) with a clustered distribution derived from the spherical harmonic analyis of the
unmasked data described above.

\begin{figure*}
\epsfig{file=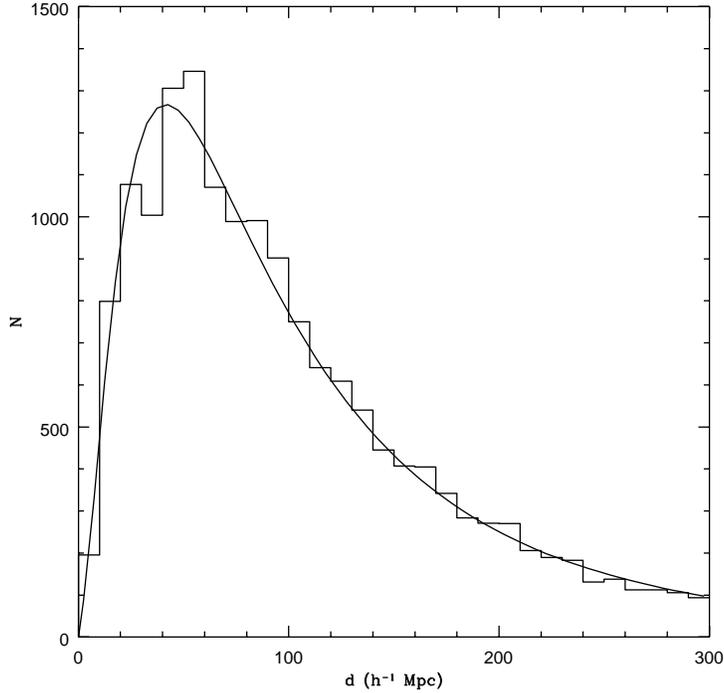,angle=0,width=10cm}
\caption{
Redshift histogram for PSCz galaxies, with assumed selection function.}
\end{figure*}

\begin{figure*}
\epsfig{file=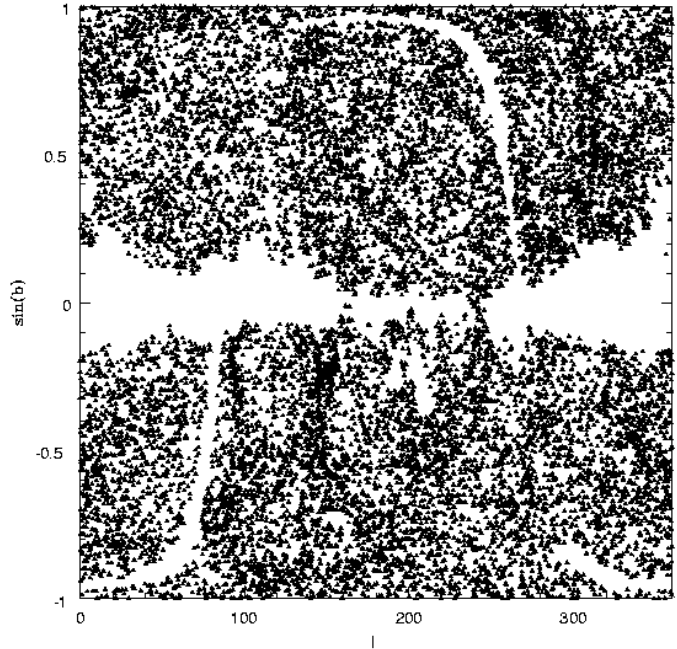,angle=0,width=10cm}
\caption{
Sky distribution of PSCz galaxies}
\end{figure*}

\begin{figure*}
\epsfig{file=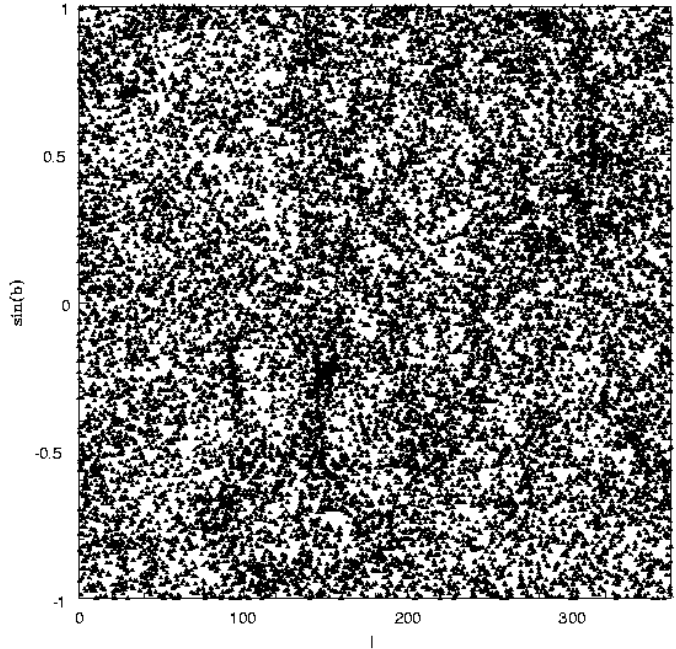,angle=0,width=10cm}
\caption{
Sky distribution of PSCz galaxies, with mask filled by sampling average sky}
\end{figure*}

\begin{figure*}
\epsfig{file=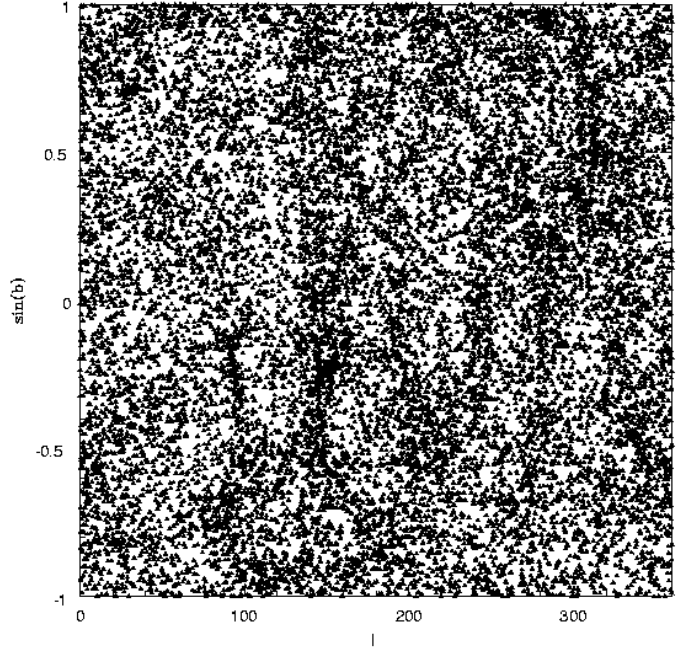,angle=0,width=10cm}
\caption{
Sky distribution of PSCz galaxies, with mask filled by sampling neighbouring bins}
\end{figure*}

\begin{figure*}
\epsfig{file=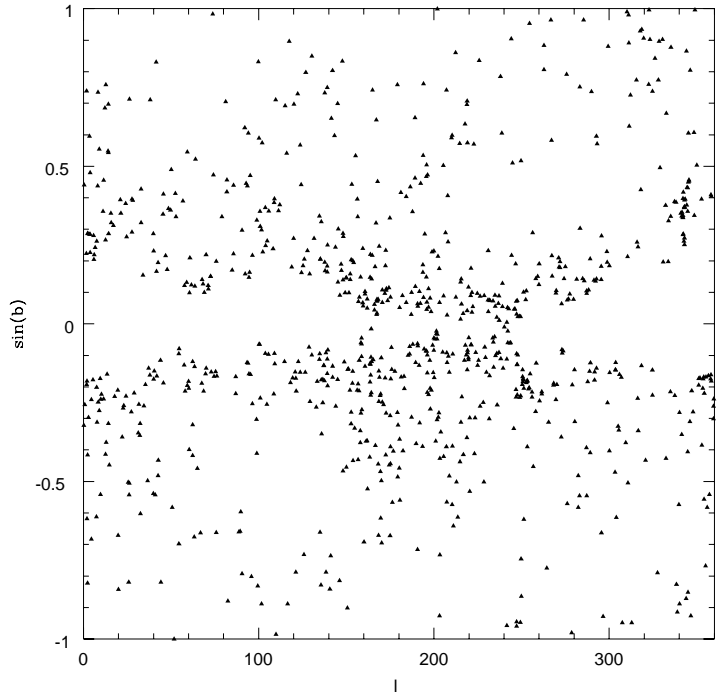,angle=0,width=10cm}
\caption{
Sky distribution of PSCz galaxies without measured redshift.}
\end{figure*}

\section{Dipole in local and CMB frames}

\begin{figure*}
\epsfig{file=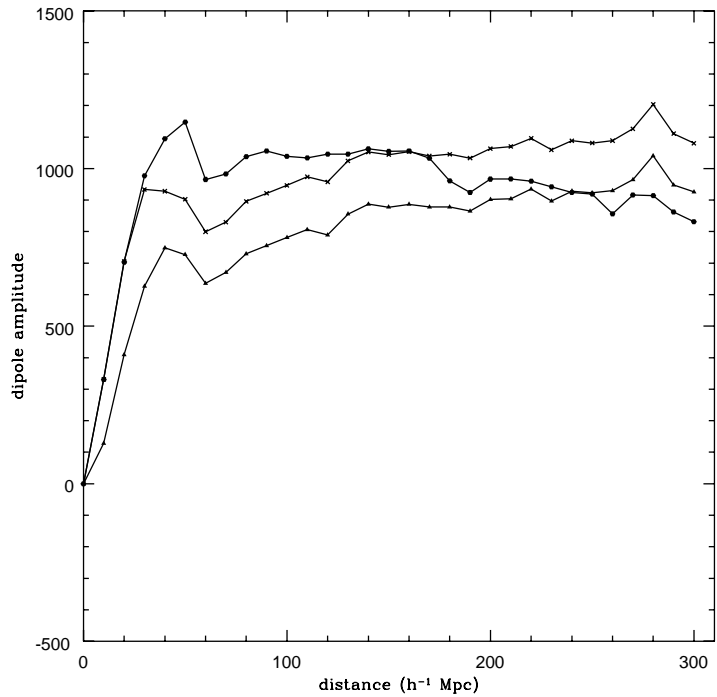,angle=0,width=10cm}
\caption{
Observed dipole amplitude in Local Group (x's) and CMB (filled triangles) frames, 
and with crude flow model (see text - filled circles).}
\end{figure*}

In linear theory the peculiar velocity at position $\bf{r}$ can be derived as (Peebles 1980)
\medskip

$\bf{V(r)}$ = $(H_o \beta/4 \pi) \int \delta(\bf{r'})(\bf{r}-\bf{r'})/|\bf{r}-\bf{r'}|^3$ $d^3 r'$ ,  (1)

where $\delta = \delta \rho/\rho$.

For a redshift survey of a galaxy population which samples the density field and satisfies
a universal luminosity function $\eta$(L,z), independently of $\delta(\bf{r'})$, the RHS
of eqn (1) can be evaluated as
\medskip

$ (H_o \beta/4 \pi n) \Sigma (\bf{r}-\bf{r'})/(\phi(|\bf{r}-\bf{r'}|) |\bf{r}-\bf{r'}|^3)$   ,  (2)
\medskip

where n is the average density of galaxies and $\phi(r)$ is the selection function.
We refer to $\bf{V(0)}$, the predicted velocity of the Local Group, as the cosmological
dipole.  The problem is that eqn (2) requires us to know the true distance of
galaxies, whereas we know only the observed radial velocity, which includes the
effect of the galaxy's peculiar velocity. 

We have first calculated the dipole as in Rowan-Robinson et al (1990) for several
different assumptions about the velocity field.

(1) In the Local Group frame, correcting only for the effects of Galactic 
rotation.  This would be valid if the CMB dipole was generated at distances
beyond those sampled by the PSCz data.

(2) In the CMB frame.  This would be valid if the CMB dipole were generated
very locally.

(3) In a crude model for the velocity field, in which no correction (apart from
Galactic rotation) is applied to galaxies with recession velocity V $< V_{o}$,
whereas galaxies with  V $> V_{o}$ are corrected for the CMB motion.  This
assumes that galaxies with V $< V_{o}$ move together as a block.  $V_{o}$ was
taken as 3000 km/s.

Figs 6 show the dipole amplitude, summed out to distance d, as a function of d,
 in these 3 models for the augmented
data with the masked area filled according to prescription (D) (spherical harmonics).  

Clearly different flow models can dramatically affect the dipole amplitude.

We use model (3) as the starting point for our iterative derivation of the dipole 
in real space in section 6.

\section{Cluster and void model for observed density distribution}

In order to correct the observed redshifts for peculiar motions and convert
to real space, we follow an approach similar in spirit to that of
Rowan-Robinson et al (1990).  We seek to make an analytic model of the
density field, from which the implied peculiar velocities can be calculated
and corrected.

The quality of the PSCz data allows us to be far more ambitious with this
model than Rowan-Robinson et al (1990), whose model used only 10 clusters
 (and no voids).  Here we try to use all prior knowledge about galaxy
 clustering out to 30,000 km$/$s.
Our initial input cluster lists consists of 

\begin{itemize}
\item (1) the galaxy groups with V $<$ 2500 km/s from Rowan-Robinson (1988),
which are in turn based on group lists by de Vaucouleurs (1976) and Geller
and Huchra (1983).
These are included to try to give some resolution of nearby structure.

\item (2) a compilation of Abell clusters with V $<$ 30000 km/s (Abell et al 1989, 
Postman et al 1992, Dalton et al 1994). 

\item (3) an additional list of clusters for which distance estimates
have been given in the literature (Sandage 1975, Lucey and Carter 1988, Mould
et al 1991, 1993).
\end{itemize} 

Because of the good all-sky sampling of the PSCz survey it is worthwhile to 
supplement these lists with
clusters and voids found in the PSCz data set itself as follows.  The data
binned in 100 sq deg bins as in section 3 is further binned in 40 velocity
bins, whose boundaries are chosen using the selection function to give the
same number of galaxies in each velocity bin (on average about 1 per bin).
The 413 x 40 array is then searched for bins in which the number of galaxies
exceeds a threshold (taken to be 4 galaxies in a bin) to yield a supplementary
list of clusters.  The array is also searched for voids,
which are defined to be sequences of 5 contiguous velocity bins containing
no galaxies.  A few (25) additional potential clusters which seemed to be visible in
the 3-dimensional distribution, but which were missed by this procees, were 
inserted by hand (numbered PC1000 onwards).

The complete list of clusters and voids is subject to a neighbours
search to weed out duplicates.

(a) d $<$ 25 $h^{-1}$ Mpc
\begin{itemize}
\item rationalize all cluster or group pairs within 2 $h^{-1}$ Mpc of each other
(this eliminates 6 nearby groups)
\item eliminate all PSCz-selected clusters within 4 $h^{-1}$ Mpc of known groups or
clusters
\item rationalize void pairs within 4 $h^{-1}$ Mpc of each other
\end{itemize}

(b) d $>$ 25 $h^{-1}$ Mpc
\begin{itemize}
\item rationalize all cluster pairs within 4 $h^{-1}$ Mpc of each other
(this eliminates 21 clusters)
\item eliminate all PSCz-selected clusters within 8 $h^{-1}$ Mpc of a known cluster
\item rationalize void pairs within 8 $h^{-1}$ Mpc of each other
\end{itemize}

This procedure is necessary to avoid clusters disrupting their neighbours during the iteration.
The final input cluster list contains 88 nearby groups, 854 clusters and 163 voids (cluster/void list
and parameters for final flow model available by ftp).
An iterative least-squares process now selects the PSCz galaxies within a specified distance
of the cluster or void centres and looks for infall (clusters)

$ V  =  A r^{-\alpha}$,	 (3)

(equivalent to a characterist density profile  $\rho (r) \propto r^{-1-\alpha}$) 

 or outflows (voids)

$ V  = A r$.

(valid for a region of constant below-average density)

If a void is found to be completely empty of galaxies, so A can not be estimated in this way, 
then A is set to be

A = 100 $h \beta /3$

which is the relevant value for an empty void.  

From a least-squares analysis we find  $\alpha = 0.6 \pm 0.1$ for cluster infall and we 
adopt the value $\alpha$ = 0.6.

We set a limit for the size of clusters and voids defined as follows:

\begin{itemize}
\item for nearby groups (and voids) within 25 $h^{-1}$Mpc, a maximum radius of 4 $h^{-1}$Mpc, except for
the Virgo, Fornax and Puppis clusters, and the Local Void (see below):
\item for all other clusters within 100 $h^{-1}$Mpc, a maximum radius of 16 $h^{-1}$Mpc:
\item for all other voids within 100 $h^{-1}$Mpc, a maximum radius of 8 $h^{-1}$Mpc:
\item for clusters with distances greater than 100 $h^{-1}$Mpc, a maximum radius of 32 $h^{-1}$Mpc:
\item for voids with distances greater than 100 $h^{-1}$Mpc, a maximum radius of 16 $h^{-1}$Mpc.
\end{itemize}

A maximum size has to be set to prevent clusters swallowing their neighbours.  The
increase in maximum size with distance reflects the worsening resolution of the model.

The cluster-void model is then used to predict the peculiar velocity
field and the radial distance of each PSCz galaxy is adjusted accordingly before
entering the next iteration. The triple-valued zone around each cluster is treated as follows.
For a cluster with amplitude A, a core radius $(A/100)^{1/(1+\alpha)} h^{-1}$Mpc is defined within
which infall to the cluster can overwhelm the Hubble flow.  Within this zone there is
ambiguity about which side of the cluster the galaxy lies on (in front or behind).  If the cluster model places 
the galaxy within this
distance of the centre of the cluster, the galaxy is assigned a distance equal to that of the cluster centre and
then is not included in the infall solution.
To avoid excessive feedback, the core radius is in practice damped by multiplying by 0.75.  Without this, clusters
tend to grow in size and swallow all the galaxies around them into their core, thereby disappearing from the
solution.

The local volume, d $<$ 25 h$^{-1}$ Mpc, requires especial care, as the resolution of the IRAS sample is
not adaquate for arriving at an accurate model.  It is clear that Virgo and Eridanus-Fornax are dominant 
structures in the local flow.  There is the possibility of additional significant structures behind the Galactic
plane.  For example it has been proposed that there is an important local structure in the
direction of Puppis (Lahav et al 1993) and several other groups or clusters within this volume at $|b| < 20^o$
have been identifed in the PSCz cluster searches described above.  Finally, inspection of the 3-dimensional galaxy 
distribution shows that 
the Local Void (Saunders et al 1991) occupies most of one quadrant of the local volume.  These 
structures have been treated as follows:  Virgo, Eridanus, Puppis and the other clusters/groups at
$|b| < 20^o$  were allowed to grow to a maximum size of 16 $h^{-1}$Mpc, ie they were 
assumed to be comparable to an Abell cluster.  The Local Void was
allowed to grow to a maximum size of 8 $h^{-1}$Mpc, ie was assumed to be comparable to other voids at 
d $<$ 100 $h^{-1}$Mpc .  The resulting model of the local 
flow can be compared with the results form the Least Action analysis of Sharpe et al (1999), in which
the orbits of the nearby galaxies are followed in detail.  To bring our results into agreement with those
of Sharpe et al, we had to move the centre of the Local Void from d = 25 $h^{-1}$Mpc, estimated by Saunders 
et al (1991) from analaysis of the QDOT sample, to 9 $h^{-1}$Mpc.  At this distance it exerts a dynamic
effect on our Galaxy comparable to that of the Virgo cluster.  The Puppis cluster was not found to 
have a very strong effect on our Galaxy's motion.  However two other clusters, PC1000 at (l,b) = (310,5), 
d = 14.5 $h^{-1}$, and PC1001 at (l,b) = (279,10), d = 28.4 $h^{-1}$Mpc, proved
to be important new structures, with mass comparable to Virgo (the centre of the former lies in the
masked region, and it is less prominent when the mask is filled homogeneously).  In order of the peculiar 
velocity generated at the Local Group (given in brackets in km s$^{-1}$), the 10 most significant individual 
structures are Virgo (162), the Local Void (127), PC1000 (119), PC1001 (67), AWM7 (part of Per-Pis, 62), 
A3526 (Cen, 42), A2052 (Her, 40), N2997 gp (33), A3565 (N.Cen, 31), and S805 (Pavo, 29).

The goodness of fit of the infall model for clusters can be tested by calculating the change in $\chi^2$ when 
the model is implemented, using the velocity error estimates of Tayor and Valentine (1999).  For
the spherical harmonic mask-fill, $\chi^2$ changes from 8247 to 4583 for 4111 degrees of freedom,
demonstrating that the fit to the local infall by eqn (3) is good.  

\section{Dipole in real space}
The cluster-void model is used, as described in the previous section, to convert
the PSCz data to a real space data set.  Note that this model is fully
non-linear, to within the limitations implied by the assumed 
spherical symmetry of the clusters and voids, although the data to which it is being fitted are based
on the linear assumption (eqn (1)). The initial velocity field is defined by the crude model (3) of
section 3, and is then improved in a series of iterations until a a self-consistent
model of the flow-field is obtained.  We have also iterated the value of $\beta$ in eqn (1) in order
to arrive at a fully self-consistent flow model. 

\begin{figure*}
\epsfig{file=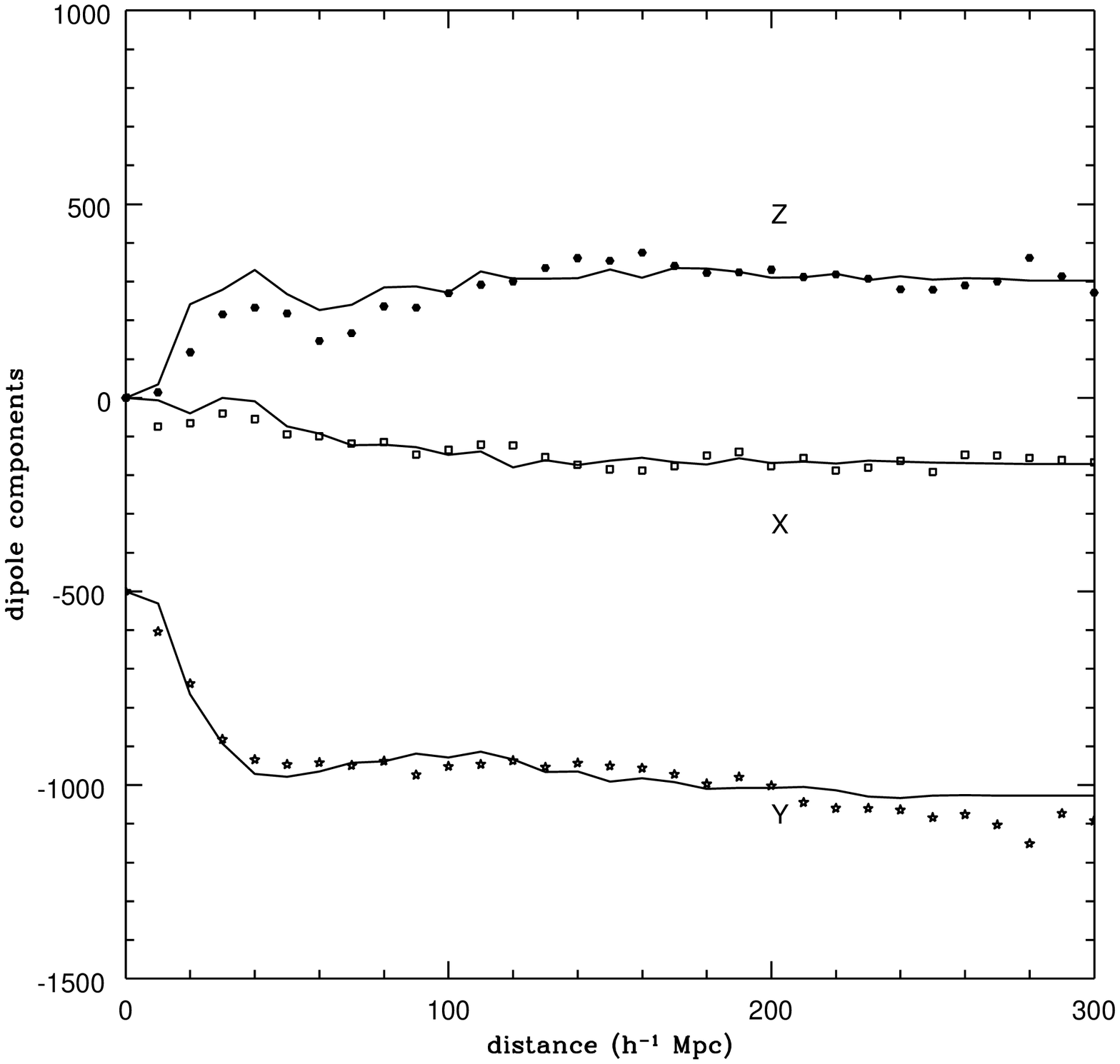,angle=0,width=10cm}
\caption{
Observed dipole components in real space, mask filled by spherical harmonics.  Solid line
shows prediction of cluster+void model.}
\end{figure*}

\begin{figure*}
\epsfig{file=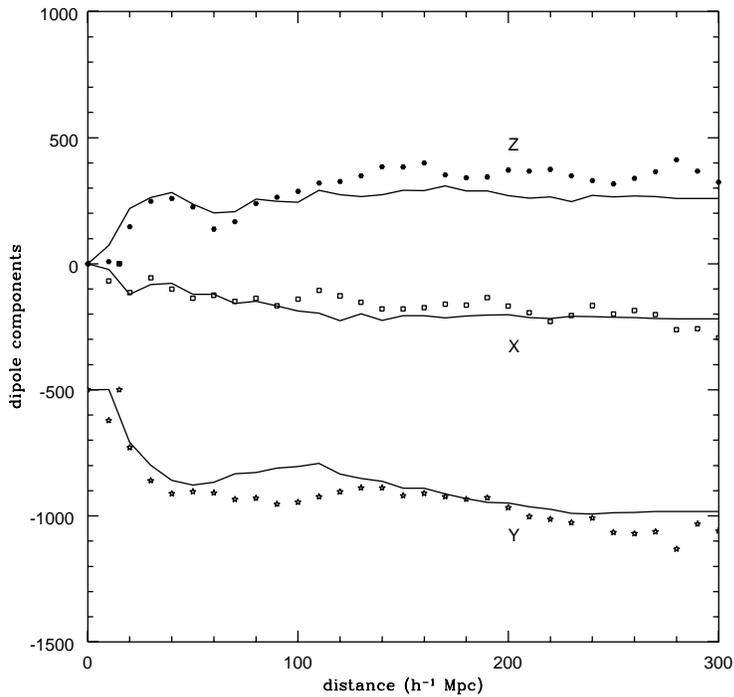,angle=0,width=10cm}
\caption{
Observed dipole components in real space, mask filled by average sky.  Solid line
shows prediction of cluster+void model.}
\end{figure*}

\begin{figure*}
\epsfig{file=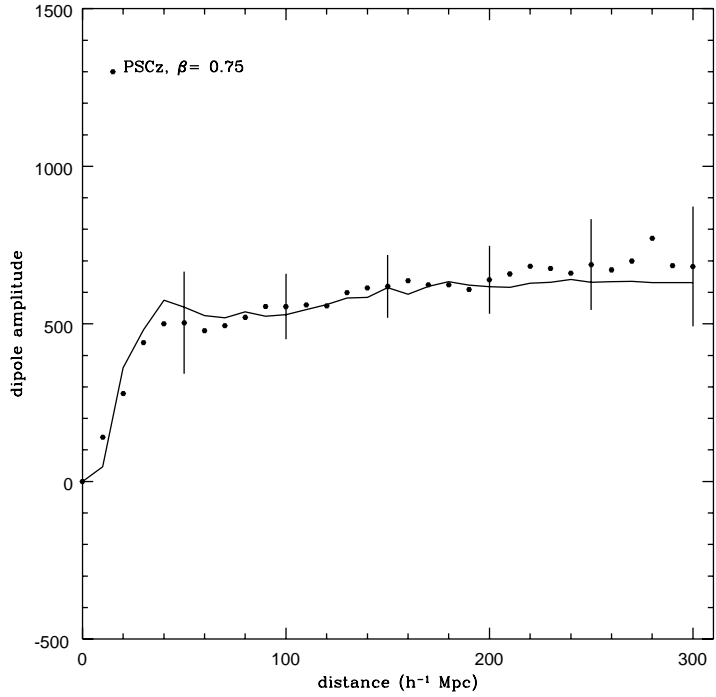,angle=0,width=10cm}
\caption{
Observed dipole amplitude in real space, mask filled by spherical harmonic
model (filled circles).  Solid line
shows prediction of cluster+void model.}
\end{figure*}
 
\begin{figure*}
\epsfig{file=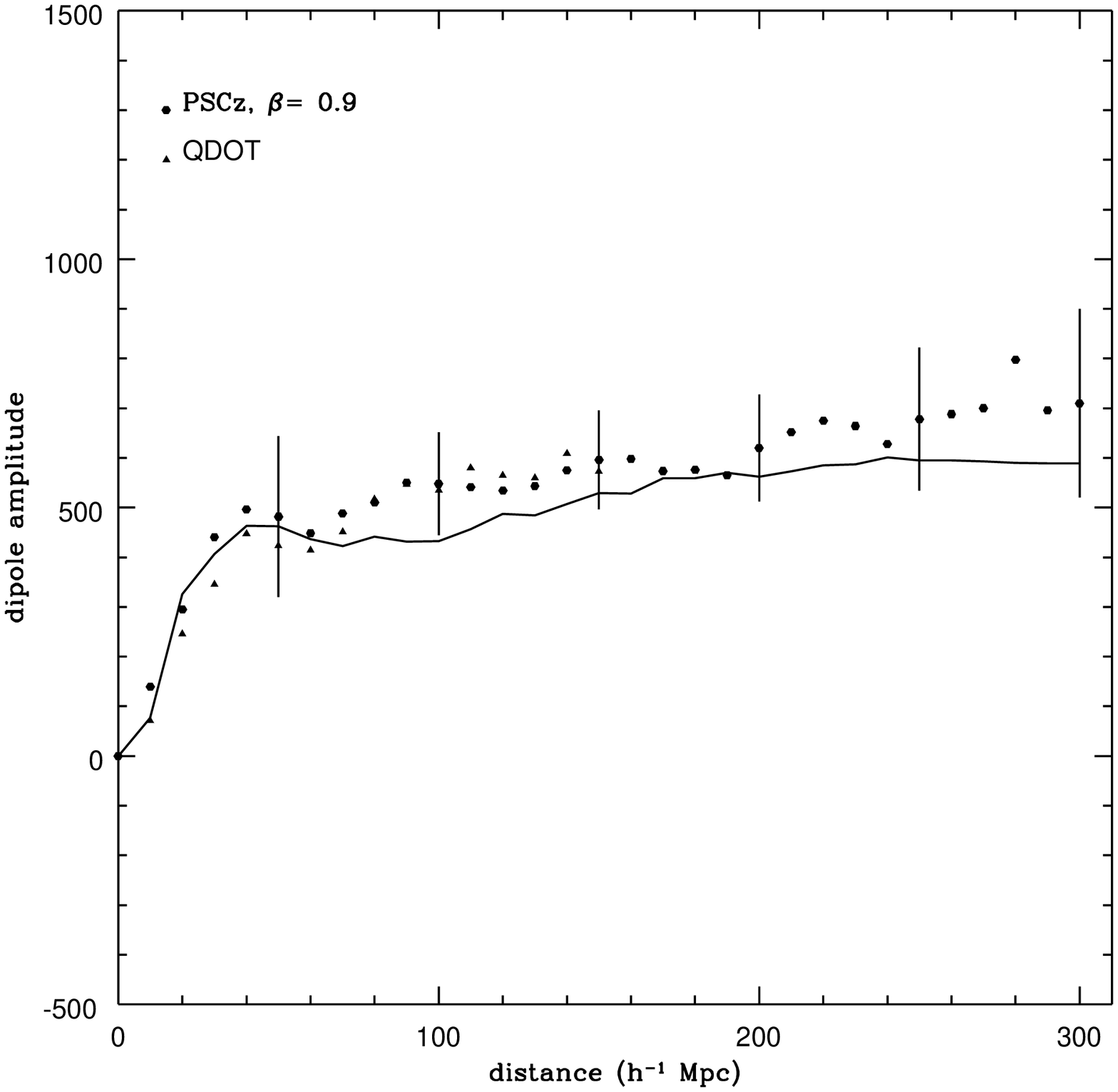,angle=0,width=10cm}
\caption{
Observed dipole amplitude in real space, mask filled by average sky (filled circles).
  Solid line
shows prediction of cluster+void model. Results from QDOT analysis (Lawrence et al 1999)
shown as filled triangles.}
\end{figure*}

Fig 7,8 shows the components of the dipole in real space, with the predictions
of the cluster-void model, for the two different mask treatments (C) and (D).  The fit of the
cluster/void model to the PSCz dipole components is excellent.
  In calculating the dipole from IRAS 
galaxies we have excluded the contribution of galaxies within 4 $h^{-1}$ Mpc.  Similarly
in computing the total dipole contribution of the cluster and group model we have excluded
groups within the same volume.  The lack of adaquate sampling in the local volume could
result in a constant offset vectors between
each of the IRAS dipole, the cluster+void model dipole, and the true dipole.  The CMB dipole
also has to include the not very well determined vector of the sun's motion with respect to
the Local Group of galaxies.  We have assumed that the sun's velocity with respect to
the Local Group is 300 km$/$s towards (l,b) = (90,0), consistent with the values
measured by Yahil et al (1977).  The offset found between the PSCz dipole and the CMB direction
is 21.7$^o$ at 150 $h^{-1}$Mpc, 18.0$^o$ at 200 $h^{-1}$Mpc, 15.5$^o$ at 250 $h^{-1}$Mpc, and 
13.4$^o$ at 300 $h^{-1}$Mpc.
Strauss et al (1992) analyzed the expected offset in a range of models using simulations and compared these 
predictions with the data for the 1.2 Jy sample.  Although the misalignment between the IRAS
and CMB dipoles has been advocated as a test of cosmological parameters (Juskiewicz et al 1990,
Lahav et al 1990, Strauss et al 1992), the problem is that density structure in the masked region could have 
a strong effect on the x and y components (especially the x component).  It is noticable
that the component which shows the greatest discrepancy with the CMB dipole is the x-component.
A cluster behind the Galactic centre at the distance of, and with twice the mass of, the
Centaurus cluster (A3526) would bring the dipoles to alignment within 5$^o$.  Since such a hypothetical
object can not be rules out, there is little cosmological information in the dipole misalignment.
Other possible reasons for misalignment are that IRAS galaxies may be subtly biased with respect to
the mass distribution or that there may be non-linear corrections to eqn (1).    

Fig 9,10 shows the total dipole amplitude as a function of distance for the two assumptions about
the mask.  The uncertainties in the dipole amplitude for the PSCz sample have been estimated by 
Taylor and Valentine (1999), 
including the effects both of cosmic variance and shot noise and these are included in Figs 9, 10.
The prediction for the cluster/void model 
is also shown in each case.  There is some difference between the predicted dipole amplitudes
for the two mask-filling assumptions,
amounting to 140 $km/sec$ at 200 $h^{-1}$ Mpc for an assumed $\beta$ =1.  Thus some of the disagreements about
values of $\beta$ in different studies stem from different assumptions about
how to fill the mask.  
It is clearly worth investigating other methods of filling the masked sky.  
However the missing information can not
actually be recovered and we have to regard the different amplitudes found from the two
methods of filling the mask as indicative of the uncertainty induced by the incomplete sky
coverage of PSCz.  Also plotted in Fig 10 is the dipole amplitude from the QDOT study 
by Rowan-Robinson et al (1990, as corrected by Lawrence et al 1999), which should be comparable 
with the case where the
mask is filled by the average sky.  The agreement is excellent over the distance range in common. 

Other approaches to the problem of dynamical reconstructing include the iterative schemes of Yahil
et al (1991) and the spherical harmonic analysis with Wiener filtering of Fisher et al (1995),
Heavens and Taylor (1995).  Our method is not only stabler than the method of Yahil et al (1991),
but it also yields an analytic model of the flow throughout the survey volume.  The spherical
harmonic approach is less subject to the bias of starting from known cluster and group lists, 
entailed in our approach.  However there is a considerable cosmographic benefit in being able
to relate the flow to identifiable structures and in a subsequent paper we will demonstrate
the benefit of this for studies of the Hubble constant (and claimed anomalies like the Lauer-Postman
effect).  

Since we can be sure that some galaxy structures extend across the masked regions, the
treatment of filling the mask homogeneously is unlikely to be realistic, so in what follows 
we give results for the case where the mask is filled using spherical harmonics.

\begin{table*}
\caption{PSCz and cluster/void model dipoles as a function of depth, mask filled by average sky}
\begin{tabular}{lllllllllllll}
 & {\bf PSCz} & & & & &  & {\bf cluster} & & & & & \\
 & & & & & & & {\bf model} & & & & & \\
{\bf d($h^{-1}$ Mpc)} & {\bf $V_{x}$} & {\bf $V_{y}$} & {\bf $V_{z}$} & {\bf $V_{tot}$} & {\bf l} & {\bf b} & {\bf $V_{x}$} 
& {\bf $V_{y}$} & {\bf $V_{z}$} & {\bf $V_{tot}$} & {\bf l} & {\bf b}\\
 & & & & & & & & & & & & \\
  10. &  -68.8 & -114.7 &   43.3 &  140.6 &  239.1 &   18.0 &   -6.2 &  -31.6 &   34.4 &   47.1 &  258.9 &   46.9 \\
  20. &  -81.7 & -220.1 &  150.4 &  278.9 &  249.6 &   32.6 &  -39.8 & -265.8 &  241.2 &  361.2 &  261.5 &   41.9 \\
  30. &  -62.6 & -360.4 &  246.3 &  440.9 &  260.1 &   34.0 &   -0.4 & -392.7 &  278.5 &  481.8 &  269.9 &   35.3 \\
  40. &  -66.3 & -420.6 &  262.7 &  500.3 &  261.0 &   31.7 &   -9.5 & -470.7 &  329.9 &  575.3 &  268.8 &   35.0 \\
  50. & -105.3 & -429.4 &  241.0 &  503.5 &  256.2 &   28.6 &  -73.4 & -478.8 &  267.1 &  553.6 &  261.3 &   28.8 \\
  60. & -111.3 & -431.0 &  175.0 &  478.3 &  255.5 &   21.5 &  -92.0 & -465.2 &  226.9 &  526.3 &  258.8 &   25.5 \\
  70. & -131.9 & -434.5 &  194.5 &  494.0 &  253.1 &   23.2 & -122.8 & -442.6 &  240.3 &  518.9 &  254.5 &   27.6 \\
  80. & -126.9 & -430.4 &  264.4 &  520.8 &  253.6 &   30.5 & -120.8 & -439.3 &  285.5 &  538.1 &  254.6 &   32.0 \\
  90. & -159.0 & -463.2 &  261.3 &  555.1 &  251.1 &   28.1 & -127.8 & -418.8 &  287.4 &  524.3 &  253.1 &   33.2 \\
 100. & -149.1 & -442.8 &  299.2 &  554.8 &  251.4 &   32.6 & -147.1 & -428.7 &  271.8 &  529.0 &  251.1 &   30.9 \\
 110. & -133.9 & -438.6 &  322.0 &  560.3 &  253.0 &   35.1 & -138.4 & -413.9 &  325.8 &  545.1 &  251.5 &   36.7 \\
 120. & -135.6 & -428.9 &  328.7 &  557.1 &  252.5 &   36.2 & -179.2 & -433.7 &  307.1 &  561.3 &  247.6 &   33.2 \\
 130. & -168.4 & -445.3 &  363.4 &  598.9 &  249.3 &   37.4 & -161.5 & -465.8 &  307.8 &  581.7 &  250.9 &   31.9 \\
 140. & -186.6 & -435.6 &  391.0 &  614.3 &  246.8 &   39.5 & -173.6 & -464.7 &  308.3 &  584.5 &  249.5 &   31.8 \\
 150. & -197.2 & -443.2 &  383.8 &  618.6 &  246.0 &   38.4 & -162.9 & -491.5 &  331.8 &  615.5 &  251.7 &   32.6 \\
 160. & -200.7 & -448.6 &  405.1 &  636.8 &  245.9 &   39.5 & -155.0 & -481.9 &  310.7 &  594.5 &  252.2 &   31.5 \\
 170. & -189.3 & -464.5 &  370.7 &  623.7 &  247.8 &   36.5 & -165.8 & -492.8 &  334.5 &  618.7 &  251.4 &   32.7 \\
 180. & -161.9 & -489.3 &  352.4 &  624.3 &  251.7 &   34.4 & -172.8 & -509.4 &  333.8 &  633.5 &  251.3 &   31.8 \\
 190. & -152.5 & -471.7 &  354.4 &  609.4 &  252.1 &   35.6 & -156.0 & -506.8 &  325.6 &  622.8 &  252.9 &   31.5 \\
 200. & -188.9 & -494.0 &  360.9 &  640.3 &  249.1 &   34.3 & -168.2 & -507.1 &  309.8 &  618.1 &  251.7 &   30.1 \\
 210. & -168.1 & -536.8 &  342.1 &  658.4 &  252.6 &   31.3 & -164.9 & -504.9 &  311.2 &  616.1 &  251.9 &   30.3 \\
 220. & -200.2 & -552.1 &  348.5 &  682.9 &  250.1 &   30.7 & -169.8 & -513.2 &  319.9 &  628.6 &  251.7 &   30.6 \\
 230. & -193.3 & -552.4 &  338.1 &  675.9 &  250.7 &   30.0 & -161.9 & -529.6 &  303.4 &  632.0 &  253.0 &   28.7 \\
 240. & -175.3 & -556.5 &  310.9 &  661.1 &  252.5 &   28.0 & -165.1 & -533.1 &  313.9 &  640.8 &  252.8 &   29.3 \\
 250. & -203.4 & -578.4 &  312.0 &  687.9 &  250.6 &   27.0 & -167.0 & -527.9 &  304.6 &  632.4 &  252.5 &   28.8 \\
 260. & -160.1 & -567.9 &  320.4 &  671.4 &  254.3 &   28.5 & -168.6 & -526.5 &  309.1 &  633.9 &  252.3 &   29.2 \\
 270. & -162.3 & -594.9 &  330.0 &  699.3 &  254.7 &   28.2 & -170.0 & -527.3 &  308.1 &  634.5 &  252.2 &   29.1 \\
 280. & -167.7 & -643.0 &  391.7 &  771.4 &  255.4 &   30.5 & -171.0 & -526.8 &  302.0 &  631.3 &  252.0 &   28.6 \\
 290. & -173.7 & -565.5 &  344.4 &  684.5 &  252.9 &   30.2 & -171.0 & -526.8 &  302.0 &  631.3 &  252.0 &   28.6 \\
 300. & -179.1 & -584.3 &  302.1 &  681.7 &  253.0 &   26.3 & -171.0 & -526.8 &  302.0 &  631.3 &  252.0 &   28.6 \\
 & & & & & & & & & & & &\\
 & & & & & {\bf CMB} & &  {\bf -25.2} & {\bf -545.4} & {\bf 276.5} & {\bf 612 $\pm$ 22} & {\bf 268 $\pm$ 3} 
 & {\bf 27 $\pm$ 3}\\
\end{tabular}
\end{table*}
\medskip

\section{Convergence of the dipole and the value of $\beta$}
Previous studies of the IRAS dipole have been limited to d $\leq$ 150 $h^{-1}$Mpc.  Our larger sample 
allows us to investigate the convergence of the dipole to a significantly greater depth.  Although
there are massive structures in the distance range 150-300 $h^{-1}$Mpc, 
it appears that the impact of these structures on the motion of the Local Group is small.  The change in dipole
amplitude from 150 to 300 $h^{-1}$Mpc is no greater than 70 km$/$s.  The Shapley cluster concentration,
for example, proposed by Raychaudhury (1989), Scaramella et al (1989), and Plionis and Valdarnini (1991) 
as a major contributor to the Local Group's motion, contributes only 21 km$/$s.  The results found here for
d $>$ 150 $h^{-1}$Mpc appear to differ from those of Strauss et al (1992) based on the 1.2 Jy sample, who saw a 
steep increase in dipole amplitude between 150 and 200 $h^{-1}$Mpc.  However this can probably be attributed to
the increased shot noise associated with the smaller sample. 

Vittorio and Juskiewicz (1987) raised the question of whether the calculation of eqn (1) can
in fact be a convergent process and this has been discussed in many subsequent papers
(eg Juskiewicz et al (1990), Lahav et al (1990), Peacock (1992) and Strauss
et al (1992)).   The growth of the dipole amplitude at large distances clearly depends on the
spectrum of density perturbations on large scales.  The fact that we empirically find convergence 
(in the sense that the dipole amplitude changes by no more than 10$\%$) over
a range of distances corresponding to an increase in volume of a factor of 8 is itself 
a significant constraint on the large scale spectrum.  Peacock (1992) emphasizes that plateaus
of apparent convergence can appear in simulations even when the the dipole amplitude is
far from the asymptotic value.  However the fact that we are seeing convergence over scales
approaching those on which the microwave background radiation is known to be extremely smooth
does not leave much scope for significant jumps in the dipole amplitude at larger distances 
than those surveyed here.

The IRAS selection at 60 $\mu$m undersamples the elliptical galaxy population, and hence
the dense cores of rich clusters.  However these cores represent only a few percent of the total
masses of the clusters characterized here, which tend to be of supercluster dimensions.
Strauss et al (1992) have shown that correction for this undersampling of cluster cores
changes the dipole amplitude by at most a few percent.

Table 1 gives the dipole components, total amplitude and direction as a function of distance,
together with the corresponding quantities for the cluster/void model.  The amplitudes and 
directions can be compared with the results of COBE (Kogut et al 1993, Fixsen et al 1996), which can be
combined with the estimate of the velocity of the sun relative to the Local Group
 to give the values shown in the last line of the Table.  From the values at 200 $h^{-1}$Mpc,
 beyond which we find no evidence for a significant contribution to the dipole, we conclude that
$\beta$ = 0.75, with a $1-\sigma$ range of 0.67-0.86 and a $2-\sigma$ range of 0.60-1.01.
The uncertainty in $\beta$, derived from the analysis of Taylor and Valentine (1999), includes
the effects of shot noise and cosmic variance, but does not include the uncertainties associated
with the mask-filling assumptions.  From repeated realizations of the average sky mask-filling, we estimate
the statistical contribution to the uncertainty in $\beta$ to be only $\pm$0.05, for a given assumption 
about how the mask should be filled.  However changing from the spherical harmonic mask-fill
to an average sky mask-fill resulted in a shift of $\beta$ by 20$\%$, which can be taken as an indication
of the maximum additional systematic uncertainty associated with the mask. 
For b = 1, the corresponding value of
$\Omega_{o}$ = 0.62, with a $1-\sigma$ range of 0.51-0.78 and a with a $2-\sigma$ range of 
0.43-1.02.  Values of $\Omega_o$ outside this range would probably require pathological behaviour of
the density fluctuation spectrum.  The present work can not decisively choose between current 
popular models with 
$\Omega_o = 1, \Lambda = 0$ and $\Omega_o = 0.3, \Lambda = 0.7$, though the former is slightly preferred.

Our value of $\beta$ can be compared with those derived from other studies of the PSCz sample: 0.7 +0.35, -0.2
from a likelihood analysis of the Local Group acceleration (Schmoldt et al 1999), 0.58 $\pm$ 0.26 from 
spherical harmonic analysis of the redshift space distortion (Tadros et al 1999), 
0.6 +0.22, -0.15 from a study of the cosmic velocity field (Branchini et al 1999).  Our result is clearly 
consistent with all of these, which is impressive given the very different approach to modelling
the density field used in these different studies.  Our results are also consistent with the earlier 
results using previous IRAS samples summarized by
Rowan-Robinson (1997, summary value 0.85 $\pm$ 0.15), and with subsequent results 
from da Costa et al (1998)
and Sigad et al (1998) (0.6 $\pm$ 0.1 and 0.89 $\pm$ 0.12, respectively).  

\section{Cluster masses}

The mass of each cluster (or mass-deficit of each void) can be calculated from

$M_{cl} = (4 \pi/3) \int \rho' r'^2 dr'$ 

= 3 $(\Omega_o/\beta) H_o (A r_{cl}^{1.4})/2G$ 

= 3.6 x 10$^{10} A r_{cl}^{1.4} h M_{\odot}$  . (3)

The results are given in Tables 2-12 for each cluster detected with $A/\sigma_A >$ 1.0.  Fig 11 shows
the distribution of cluster masses with distance.  
The loss of resolution with increasing distance
means that only very large structures can be detected at large distances.
The masses in the Hydra-Centaurus, Pavo-Indus,
Perseus-Pisces-Cetus, Coma-A1367, Hercules and Shapley supercluster complexes are respectively 7.1, 4.4, 
11.8, 6.5, 34.2, 16.5 x 10$^{15} M_{\odot}$ compared
to the Virgo cluster mass of 1.3x10$^{15} M_{\odot}$.  Our cluster masses our generally 
substantially larger than the estimates given for the region within an Abell radius, since
they include the extensive halo of galaxies extending out to tens of Mpc around each cluster.  

Figure 12 gives a histogram of the number of clusters and groups in the input list (dotted histogram),
together with those detected in the PSCz sample (solid, shaded histogram).  Out to d = 150 $h^{-1}$ Mpc,
about 60 $\%$ of clusters are detected, but beyond this distance there is a steady fall in
the percentage of clusters detected, reflecting the selection function of Fig 1.

Figure 13 shows the average density fluctuation

$ \delta \rho / \rho = 3 M / 4 \pi \rho_o r_{cl}^3$

as a function of cluster radius $r_{cl}$.  If we take the median value as indicative of
$ <\delta \rho / \rho(r_{cl})$, then we find

$ <\delta \rho / \rho> \propto r^{-0.75}$, for 2 $\leq$ d $\leq$ 32 $h^{-1}$ Mpc,

comparable with the findings of Sutherland et al (1999).

\begin{figure*}
\epsfig{file=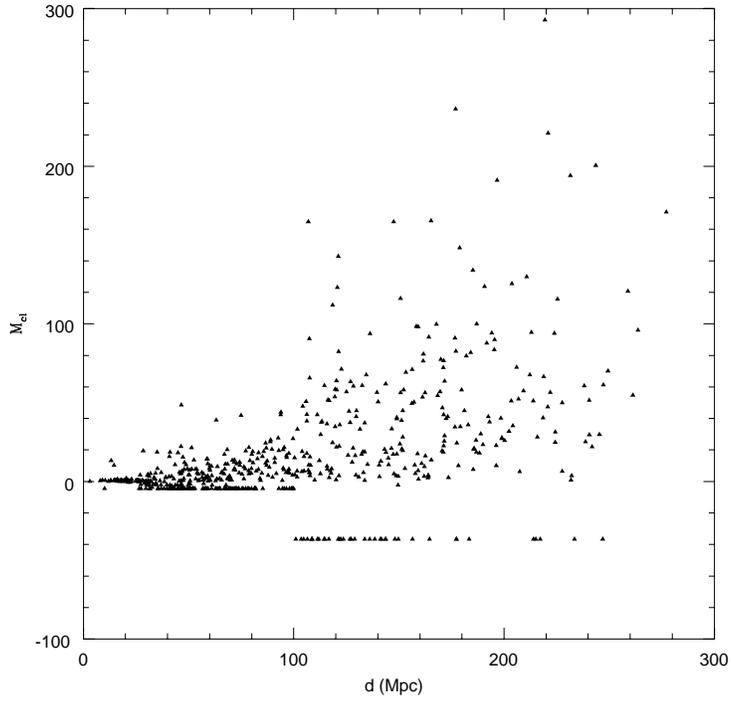,angle=0,width=10cm}
\caption{
Masses of clusters and voids (in units of $10^{14} M_{\odot}$) as a function of distance.}
\end{figure*}

\begin{figure*}
\epsfig{file=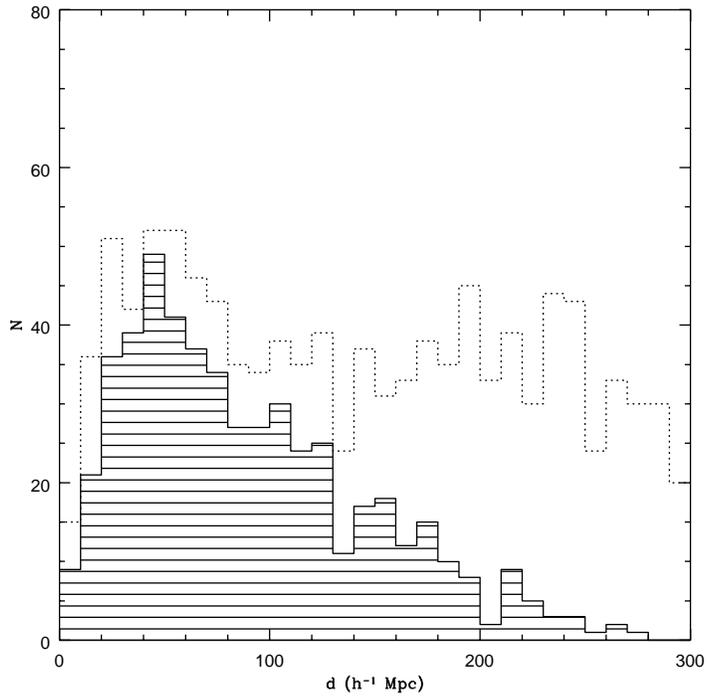,angle=0,width=10cm}
\caption{
Histogram of clusters in input catalogue (broken curve) and those detected as structures in PSCz (shaded).}
\end{figure*}

\begin{figure*}
\epsfig{file=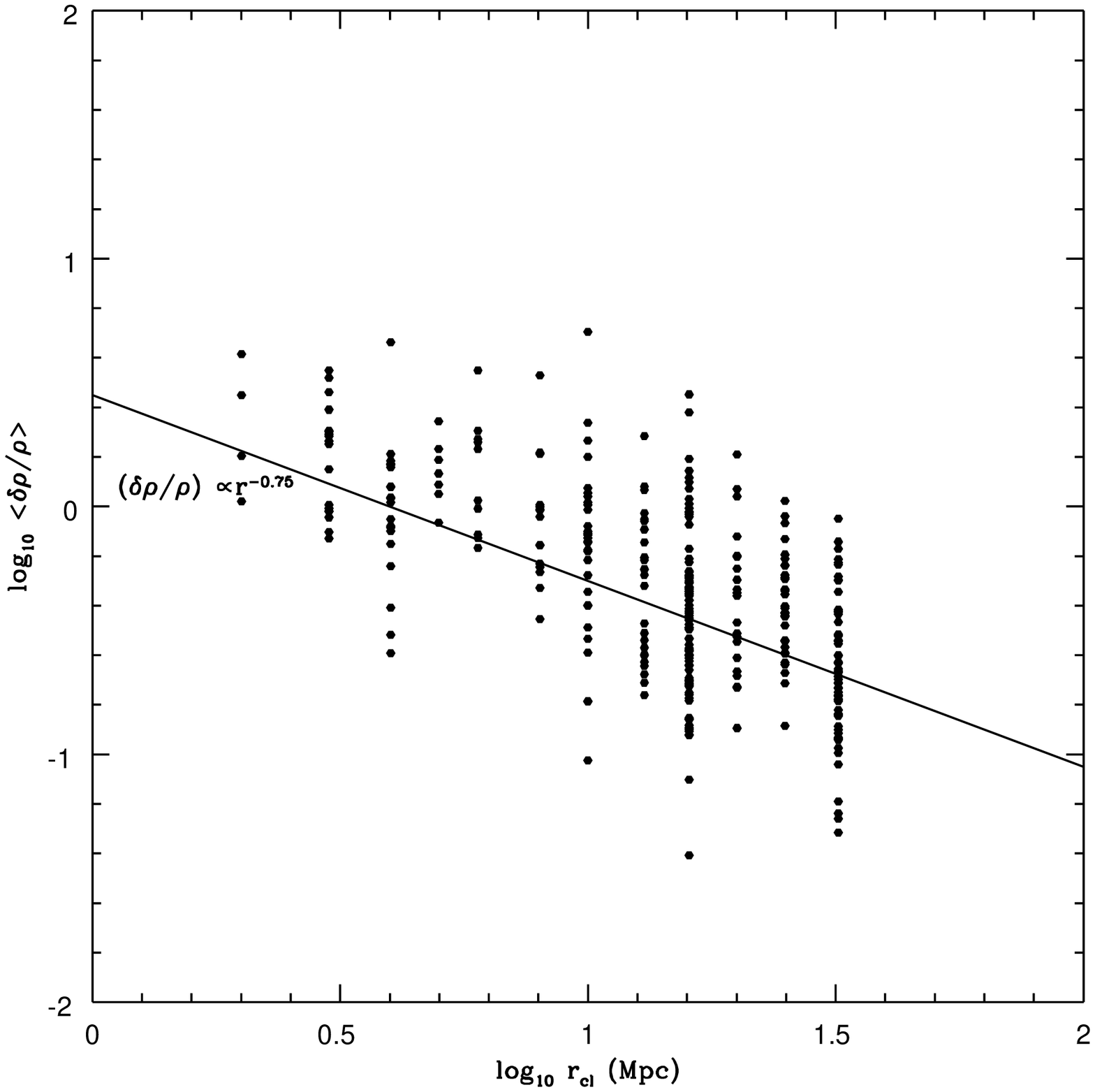,angle=0,width=10cm}
\caption{
The average value of $\delta \rho/\rho$ in a cluster versus cluster radii.  The solid line corresponds to
$\delta \rho/\rho \propto r^{-0.75}$, corresponding to $P(k) \propto k^{-1.5}$, in agreement with the power spectrum
derived by Sutherland et al (1999).}
\end{figure*}

\section{The flow field predicted by IRAS}

Fig 14 shows the distribution in Supergalactic coordinates (z-axis towards (l,b) = (47, 6), 
x-axis towards (l,b) = (137, 0)) of galaxies within 22.5$^o$ of the Supergalactic plane,
with the direction of their predicted flow (eqn 1) indicated.  The strong concentrations towards Virgo,
Hydra-Centaurus and Perseus-Pisces are clearly seen.  Fig 15 shows the corresponding
distribution for cluster centres, with the velocities predicted by the cluster-void model.
A detailed discussion of the flow field derived from PSCz is given by Saunders et al (1999).

In a later paper we will use the peculiar velocities of the clusters to study the Hubble diagram
for those clusters for which a distance is known and will give an analysis if the Lauer and Postman
result.  

\begin{figure*}
\epsfig{file=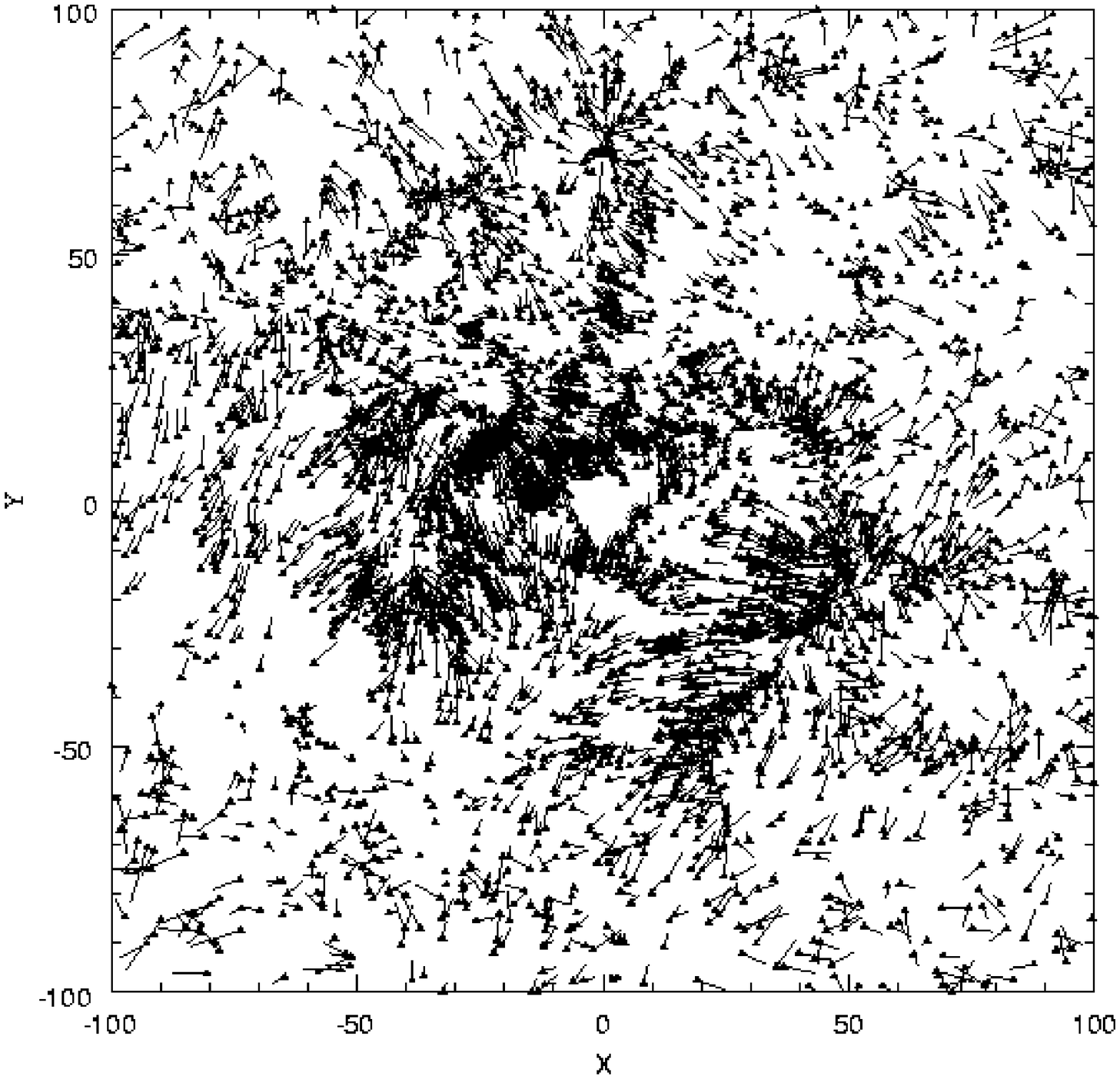,angle=0,width=10cm}
\caption{
Galaxies within 22.5$^o$ of the Supergalactic plane, with the direction of 
their predicted motion, in the CMB frame,
indicated.}
\end{figure*}

\begin{figure*}
\epsfig{file=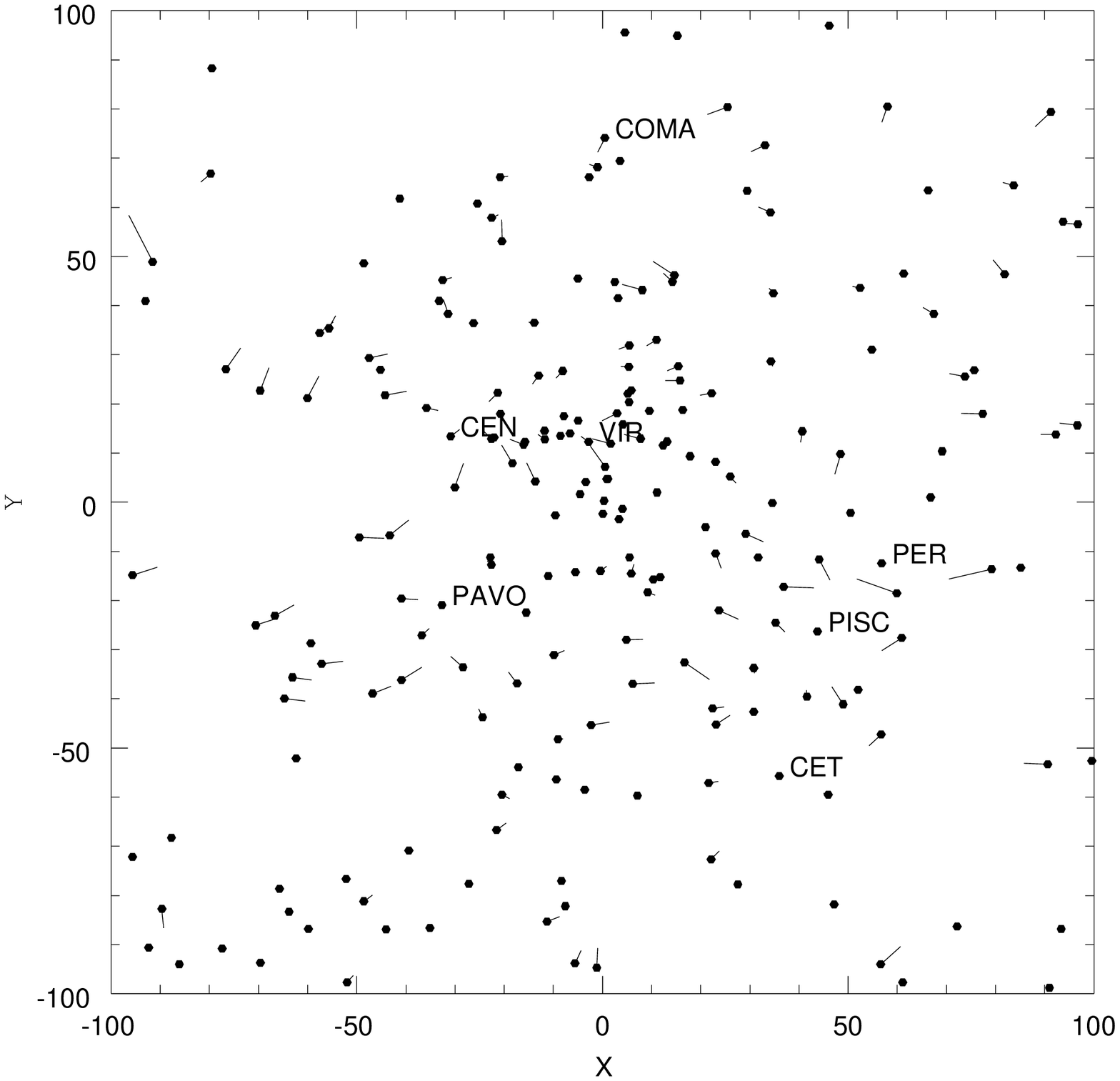,angle=0,width=10cm}
\caption{
Clusters within 22.5$^o$ of the Supergalactic plane, with the direction of their 
predicted motion
indicated.}
\end{figure*}

\section{Summary}

(1) We have investigated the convergence of the cosmological dipole using the new PSCz IRAS galaxy
redshift survey.

(2) The amplitude of the calculated dipole depends on the assumption used to fill the masked
area of the sky and this accounts for some of the differences between previous results.

(3) The PSCz identifications appear to be substantially complete to d = 300 $h^{-1}$ Mpc for areas
of the sky with I(100) $>$ 12.5 MJy$/$sr.  Between I(10) = 12.5 and 25 MJy$/$sr, the completeness limit
declines to 150 $h^{-1}$ Mpc.  This effect was corrected by adding additional sources in this zone based 
on a spherical harmonic analysis of the unmasked sky.

(4) The dipole appears to have converged by 200 $h^{-1}$ Mpc.  Between 200 and 300 $h^{-1}$ Mpc,
the additional contribution to the dipole amplitude is estimated to be $\leq$ 40 km$/$s.  Rather
special and pathological assumptions about the power spectrum of density fluctuations would be
required to achieve consistency with the present data and the CMB fluctuations, yet result in 
a very different asymptotic dipole amplitude.

(5) The direction of the dipole calculated from IRAS galaxies (the direction in which the Local
Group is being pulled) is 13$^o$ away from the CMB dipole (the direction in which the Local Group
is moving), the main uncertainty in
direction being associated with the masked area behind the Galactic plane.  The improbability
that further major contributions to the dipole amplitude will come from volumes larger
than those surveyed here means that the question of the origin of the CMB dipole is
essentially resolved.  

(6) The correction of the observed (heliocentric) velocities to real space distances is performed
using an analytic model of the flow field involving 842 clusters and 163 voids, with all Abell 
clusters within 300 $h^{-1}$ Mpc involved in the solution.  About 60 $\%$ of clusters with
d $<$ 150 $h^{-1}$ Mpc are detected as significant dynamical objects.  Inferred masses for Abell
clusters are in the range 1 - 300 x 10$^{14} M_o$.  The Local Void and two new nearby (d $<$ 30 Mpc)
clusters identified close to the Galactic plane at (l,b) = (310,5), (279,10) have a major
effect on the Local Group motion.  It will be interesting to see if redshift surveys within
the masked area of the Galactic plane can confirm the scale of these structures.

(7)  When the mask is filled with a Poissonian distribution of sources typical of the average 
unmasked sky, a value for $\beta$ = 0.90 $\pm$ 0.15 is found, in excellent agreement with
the results found from QDOT (Rowan-Robinson et al 1990, Lawrence et al 1999).  For a
spherical harmonic mask-fill, which is probably the most realistic assumption to make
about how the mask should be filled, we find the value of $\beta$ = 0.75 +0.11,-0.08.
For b =1 this corresponds to $\Omega_o$ = 0.62, with a 2-$\sigma$ range of 0.43-1.02.
Alternatively $\Omega_o$ = 1 requires a bias factor b = 1.33 $\pm$ 0.17.  The maximum
additional systematic uncertainty associated with our ignorance of the masked region 
is estimated to be $\pm$20$\%$.  

\begin{table*}
\caption{Groups and clusters with $V_{obs} < 2500 km/s$}
\begin{tabular}{lllllllllll}
{\bf name} & {\bf l} & {\bf b} & {\bf $V_{obs}$} & {\bf $V_{CMB}$} & {\bf A} & {\bf $r_{cl}$} & {\bf $S/N$} 
& {\bf $V_{LG}$} & {\bf $M_{cl} (10^{12}) M_{\odot}$} & {\bf other name} \\
 & & & & & & & & & & \\
R1 N134    &  21.8 & -86.1 &  1609.0 &  1462.6 &  492.4 &  3. &  4.7 &   11.6 &      89.3 &                  \\
R3 N448    & 136.6 & -63.9 &  1748.0 &  1309.4 &  305.7 &  4. &  2.6 &   12.4 &      76.6 &                  \\
R6 CETII   & 152.6 & -67.7 &  1827.0 &  1653.0 &  203.5 &  4. &  4.9 &    5.2 &      51.0 &                  \\
R8 CETI    & 169.0 & -55.3 &  1433.0 &  1050.2 &  134.5 &  2. &  3.7 &    3.2 &      12.8 & R10              \\
R11 N1248  & 180.1 & -46.5 &  2088.0 &  1943.1 &  306.8 &  4. &  3.1 &    5.7 &      76.9 &                  \\
R12 ERIDAN & 207.3 & -51.8 &  1549.0 &  1619.9 &  441.9 &  5. &  6.8 &   16.2 &     153.1 &                  \\
\end{tabular}
\end{table*}
\medskip 
 
 \begin{table*}
\caption{PSCz groups with $V_{obs} < 2500 km/s$}
\begin{tabular}{lllllllllll}
{\bf name} & {\bf l} & {\bf b} & {\bf $V_{obs}$} & {\bf $V_{CMB}$} & {\bf A} & {\bf $r_{cl}$} & {\bf $S/N$} 
& {\bf $V_{LG}$} & {\bf $M_{cl} (10^{12}) M_{\odot}$} & {\bf other name} \\
 & & & & & & & & & & \\                
     PC190 & 210.0 & -16.0 &  1906.9 &  2295.6 &  405.0 &  5. &  4.6 &    7.4 &     140.3 &                  \\
     PC222 & 245.0 &   0.0 &  2218.1 &  3075.0 &  495.9 & 10. &  3.4 &   13.3 &     453.4 &                  \\
     PC291 & 248.8 &  20.0 &  1807.2 &  2535.2 &  375.5 &  4. &  3.5 &    4.1 &      94.1 &                  \\
     PC315 & 249.7 &  30.0 &  2246.3 &  3066.3 &  273.5 &  3. &  3.7 &    1.5 &      49.6 &                  \\
     PC322 & 307.7 &  30.0 &  2284.4 &  2637.4 &  594.5 &  3. & 13.3 &    3.1 &      78.1 &                  \\
     PC328 & 319.4 &  30.0 &  2320.9 &  2697.3 &  202.2 &  3. & 13.8 &    1.0 &      26.5 &                  \\
     PC364 & 353.6 &  40.0 &  2044.4 &  2588.7 &  264.8 &  4. &  6.6 &    2.8 &      66.4 &                  \\
     PC388 & 305.2 &  50.0 &  1422.4 &  1901.0 &  162.0 &  3. &  8.6 &    1.6 &      21.3 &                  \\
     PC415 & 290.0 &  60.0 &  2349.0 &  2959.7 &  110.6 &  3. &  3.2 &    0.6 &      20.1 &                  \\
     PC436 & 285.0 &  70.0 &  2280.0 &  2836.5 &  151.1 &  3. &  7.7 &    0.9 &      27.4 &                  \\
    PC1000 & 310.0 &   5.0 &  1255.2 &  1451.3 &  592.8 & 16. &  5.9 &  119.1 &    1035.1 &                  \\
    PC1001 & 279.0 &  10.0 &  2207.2 &  2839.5 & 1112.1 & 16. & 14.7 &   66.9 &    1941.8 &                  \\
\end{tabular}
\end{table*}
\medskip
    
\begin{table*}
\caption{Abell and other clusters with $V_{obs} = 2500-15000 km/s$}
\begin{tabular}{lllllllllll}
{\bf name} & {\bf l} & {\bf b} & {\bf $V_{obs}$} & {\bf $V_{CMB}$} & {\bf A} & {\bf $r_{cl}$} & {\bf $S/N$} 
& {\bf $V_{LG}$} & {\bf $M_{cl} (10^{12}) M_{\odot}$} & {\bf other name} \\
 & & & & & & & & & & \\                 
  A2731    & 313.9 & -59.3 &  9354.0 &  9531.8 &  817.7 & 10. &  2.8 &    2.3 &     747.7 &                  \\
  N80      & 114.0 & -40.0 &  6268.0 &  5912.4 &  541.4 &  8. &  5.5 &    2.8 &     358.2 & GH3              \\
  N128     & 112.0 & -60.0 &  4657.0 &  3703.8 &  329.0 & 16. &  2.4 &   11.6 &     574.5 & GH6              \\
  N194     & 117.0 & -60.0 &  5298.0 &  4789.5 & 1600.2 &  6. &  8.4 &    9.3 &     766.2 &                  \\
  A76      & 117.8 & -56.0 & 11990.0 & 11580.9 & 1098.6 & 25. &  1.8 &    7.6 &    3663.3 &                  \\
  S109     & 284.7 & -86.0 &  9473.0 &  9494.2 & 1154.6 & 16. &  4.4 &    6.2 &    2016.1 &                  \\
\end{tabular}
\end{table*}
\medskip
  
  \begin{table*}
\caption{Abell and other clusters with $V_{obs} = 2500-15000 km/s$ (cont.)}
\begin{tabular}{lllllllllll}
{\bf name} & {\bf l} & {\bf b} & {\bf $V_{obs}$} & {\bf $V_{CMB}$} & {\bf A} & {\bf $r_{cl}$} & {\bf $S/N$} 
& {\bf $V_{LG}$} & {\bf $M_{cl} (10^{12}) M_{\odot}$} & {\bf other name} \\
 & & & & & & & & & & \\    
  N5056    &  67.1 &  81.7 &  6865.0 &  7167.4 &  965.4 & 13. &  1.8 &    6.6 &    1219.8 & GH119            \\
  A1736    & 312.6 &  35.1 & 13639.0 & 14755.0 & 4944.0 & 25. &  1.6 &   21.0 &   16485.8 & SHAPLEY          \\
  A3565    & 313.5 &  28.0 &  3376.0 &  4114.1 & 1069.2 & 16. & 12.3 &   30.6 &    1866.9 & IC4296,N.CEN     \\
  S740     & 314.1 &  23.6 & 10073.0 & 10172.2 &  720.3 & 32. &  1.9 &    8.9 &    3319.1 &                  \\
 A3571     & 316.3 &  28.6 & 11518.0 & 12506.1 & 3263.3 & 16. &  5.7 &   10.1 &    5698.1 & A3572            \\
 A3575     & 317.5 &  28.3 & 10802.0 & 10404.5 & 1514.9 & 13. &  2.9 &    4.9 &    1914.1 &                  \\
\end{tabular}
\end{table*}
\medskip
 
 \begin{table*}
\caption{Abell clusters with $V_{obs} =15000-30000 km/s$}
\begin{tabular}{lllllllllll}
{\bf name} & {\bf l} & {\bf b} & {\bf $V_{obs}$} & {\bf $V_{CMB}$} & {\bf A} & {\bf $r_{cl}$} & {\bf $S/N$} 
& {\bf $V_{LG}$} & {\bf $M_{cl} (10^{12}) M_{\odot}$} & {\bf other name} \\
 & & & & & & & & & & \\                 
  A85      & 115.0 & -72.1 & 15703.6 & 15835.2 & 2136.6 & 32. &  1.5 &   10.9 &    9845.5 &                  \\
  A102     & 121.5 & -61.5 & 18949.4 & 19075.7 & 2687.4 & 32. &  2.4 &    9.5 &   12383.5 &                  \\
  A104     & 122.4 & -38.3 & 24433.7 & 24361.1 & 4351.0 & 32. &  2.0 &    9.4 &   20049.4 &                  \\
  A150     & 129.6 & -49.5 & 17868.1 & 17921.2 & 1444.8 & 20. &  1.6 &    3.0 &    3486.3 &                  \\
  A151     & 142.9 & -77.6 & 16108.3 & 16539.5 & 1081.5 & 13. &  3.5 &    1.4 &    1366.5 &                  \\
  A450     & 166.0 & -24.8 & 18293.8 & 18544.5 &  610.3 & 25. &  1.7 &    1.6 &    2035.1 &                  \\
\end{tabular}
\end{table*}
\medskip
  
\begin{table*}
\caption{Dalton clusters with $V_{obs} = 15000-30000 km/s$}
\begin{tabular}{lllllllllll}
{\bf name} & {\bf l} & {\bf b} & {\bf $V_{obs}$} & {\bf $V_{CMB}$} & {\bf A} & {\bf $r_{cl}$} & {\bf $S/N$} 
& {\bf $V_{LG}$} & {\bf $M_{cl} (10^{12}) M_{\odot}$} & {\bf other name} \\
 & & & & & & & & & & \\                 
D0024-4849 & 313.9 & -68.1 & 21504.8 & 22194.7 & 1226.7 & 32. &  2.1 &    3.2 &    5652.6 &                  \\
 S210      & 291.6 & -51.2 & 21411.0 & 22095.1 & 4798.9 & 32. &  2.6 &   12.6 &   22113.3 &                  \\
 S311      & 285.7 & -46.4 & 21086.7 & 21853.4 & 2312.2 & 16. &  2.8 &    2.3 &    4037.3 &                  \\
D0320-2459 & 217.7 & -56.1 & 25980.2 & 26373.4 & 2882.3 & 25. &  3.1 &    3.8 &    9611.1 &                  \\
 A3175     & 209.8 & -48.3 & 11293.2 & 11842.1 & 1019.5 & 20. &  2.1 &    4.9 &    2460.0 &                  \\
 A3260     & 218.7 & -39.6 & 19042.8 & 19668.4 & 4147.4 & 32. &  1.3 &   13.7 &   19111.2 &                  \\
 A3809     & 356.1 & -49.6 & 18574.3 & 18901.0 &  908.4 & 25. &  1.0 &    2.4 &    3029.1 &                  \\
D2155-7206 & 318.9 & -39.5 & 20534.0 & 21073.1 & 5381.0 & 20. & 12.7 &    8.1 &   12984.2 &                  \\
D2203-4600 & 352.0 & -52.6 & 24558.2 & 24945.2 & 1522.1 & 32. &  1.5 &    3.1 &    7013.8 &                  \\
 S1166     & 310.8 & -49.8 & 21738.8 & 22388.4 & 2041.8 & 32. &  1.0 &    5.2 &    9408.6 &                  \\
\end{tabular}
\end{table*}
\medskip
 
 \begin{table*}
\caption{PSCz clusters with $V_{obs} = 2500-30000 km/s$}
\begin{tabular}{lllllllllll}
{\bf name} & {\bf l} & {\bf b} & {\bf $V_{obs}$} & {\bf $V_{CMB}$} & {\bf A} & {\bf $r_{cl}$} & {\bf $S/N$} 
& {\bf $V_{LG}$} & {\bf $M_{cl} (10^{12}) M_{\odot}$} & {\bf other name} \\
 & & & & & & & & & & \\                 
     PC 03 & 150.0 & -80.0 & 27606.0 & 27713.4 & 3709.2 & 32. &  5.1 &    6.2 &   17092.0 &                  \\
     PC 10 &  45.0 & -70.0 &  9712.6 &  9768.3 &  868.7 & 10. &  4.9 &    2.3 &     794.3 &                  \\
     PC 15 & 105.0 & -70.0 &  7939.1 &  7627.6 &  534.5 & 16. &  2.7 &    4.5 &     933.3 &                  \\
     PC 17 & 135.0 & -70.0 & 15112.6 & 15053.7 &  986.3 & 20. &  1.0 &    2.9 &    2379.9 &                  \\
     PC 21 & 165.0 & -70.0 & 15066.6 & 15338.6 & 2076.1 & 25. &  3.6 &    8.2 &    6922.8 &                  \\
     PC 27 &  90.0 & -60.0 &  6820.0 &  6353.1 &  490.3 &  5. &  8.0 &    1.2 &     169.9 &                  \\
\end{tabular}
\end{table*}
\medskip
     
 \begin{table*}
\caption{PSCz clusters with $V_{obs} = 2500-30000 km/s$ (cont.)}
\begin{tabular}{lllllllllll}
{\bf name} & {\bf l} & {\bf b} & {\bf $V_{obs}$} & {\bf $V_{CMB}$} & {\bf A} & {\bf $r_{cl}$} & {\bf $S/N$} 
& {\bf $V_{LG}$} & {\bf $M_{cl} (10^{12}) M_{\odot}$} & {\bf other name} \\
 & & & & & & & & & & \\          
     PC285 & 174.7 &  20.0 & 13366.0 & 13963.1 & 1698.0 & 25. &  3.1 &    8.1 &    5662.0 &                  \\
     PC288 & 185.3 &  20.0 & 18514.0 & 19183.2 & 3644.0 & 20. &  3.0 &    6.6 &    8792.8 &                  \\
     PC289 & 217.1 &  20.0 &  8990.0 &  9764.1 & 1023.8 & 16. &  3.1 &    5.2 &    1787.7 &                  \\
     PC293 & 259.4 &  20.0 &  8412.9 &  9100.3 &  685.8 & 16. &  3.0 &    4.0 &    1197.5 &                  \\
     PC294 & 259.4 &  20.0 & 12312.9 & 13510.5 &  621.7 & 16. &  1.4 &    1.7 &    1085.6 &                  \\
     PC302 & 312.4 &  20.0 & 14831.8 & 16122.5 & 3065.2 & 16. &  6.4 &    5.7 &    5352.2 &                  \\
\end{tabular}
\end{table*}
\medskip
    
\begin{table*}
\caption{PSCz voids}
\begin{tabular}{lllllllllll}
{\bf name} & {\bf l} & {\bf b} & {\bf $V_{obs}$} & {\bf $V_{CMB}$} & {\bf A} & {\bf $r_{cl}$} & {\bf $S/N$} 
& {\bf $V_{LG}$} & {\bf $M_{cl} (10^{12}) M_{\odot}$} & {\bf other name} \\
 & & & & & & & & & & \\                 
     PV 01 &  30.0 & -80.0 &  4776.0 &  4571.8 &  -24.8 &  8. &  0.0 &    6.1 &    -457.1 &                  \\
     PV 02 &  90.0 & -80.0 & 11342.1 & 11131.4 &  -24.8 & 16. &  0.0 &    8.2 &   -3656.9 &                  \\
     PV 03 & 150.0 & -80.0 &  3256.0 &  2907.8 &  -24.8 &  5. &  0.0 &    3.8 &    -114.3 &                  \\
     PV 04 &  15.0 & -70.0 &  5286.6 &  5012.0 &  -24.8 &  8. &  0.0 &    5.1 &    -457.1 &                  \\
     PV 05 &  45.0 & -70.0 & 12322.6 & 12183.8 &  -24.8 & 16. &  0.0 &    6.8 &   -3656.9 &                  \\
     PV 06 &  75.0 & -70.0 &  4269.1 &  3851.0 &   -1.0 &  8. &  0.0 &    0.3 &     -18.4 &                  \\
\end{tabular}
\end{table*}
\medskip

\begin{table*}
\caption{PSCz voids (cont.)}
\begin{tabular}{lllllllllll}
{\bf name} & {\bf l} & {\bf b} & {\bf $V_{obs}$} & {\bf $V_{CMB}$} & {\bf A} & {\bf $r_{cl}$} & {\bf $S/N$} 
& {\bf $V_{LG}$} & {\bf $M_{cl} (10^{12}) M_{\odot}$} & {\bf other name} \\
 & & & & & & & & & & \\      
     PV 86 & 210.9 & -10.0 &  7898.3 &  8529.2 &  -24.8 &  8. &  0.0 &    1.7 &    -457.1 &                  \\
     PV 87 & 210.9 & -10.0 & 13988.3 & 14810.7 &  -24.8 & 16. &  0.0 &    4.6 &   -3656.9 &                  \\
     PV 88 & 221.1 & -10.0 &  4385.8 &  4945.9 &  -24.8 &  8. &  0.0 &    5.2 &    -457.1 &                  \\
     PV 89 & 231.4 & -10.0 &  4689.1 &  5226.5 &  -24.8 &  8. &  0.0 &    4.6 &    -457.1 &                  \\
     PV 90 & 231.4 & -10.0 &  9669.1 & 10359.3 &  -24.8 & 16. &  0.0 &    9.5 &   -3656.9 &                  \\
     PV 91 & 262.3 & -10.0 &  6567.2 &  7173.0 &  -24.8 &  8. &  0.0 &    2.5 &    -457.1 &                  \\
\end{tabular}
\end{table*}
\medskip

\begin{table*}
\caption{PSCz voids (cont.)}
\begin{tabular}{lllllllllll}
{\bf name} & {\bf l} & {\bf b} & {\bf $V_{obs}$} & {\bf $V_{CMB}$} & {\bf A} & {\bf $r_{cl}$} & {\bf $S/N$} 
& {\bf $V_{LG}$} & {\bf $M_{cl} (10^{12}) M_{\odot}$} & {\bf other name} \\
 & & & & & & & & & & \\      
     PV157 & 121.9 &  30.0 &  9860.6 &  9795.3 &  -24.8 &  8. &  0.0 &    1.3 &    -457.1 &                  \\
     PV158 & 133.5 &  30.0 &  7828.5 &  7809.5 &  -24.8 &  8. &  0.0 &    2.1 &    -457.1 &                  \\
     PV159 & 214.8 &  30.0 &  6341.7 &  6928.8 &  -24.8 &  8. &  0.0 &    2.6 &    -457.1 &                  \\
     PV160 & 261.3 &  30.0 & 15293.2 & 16460.6 &  -24.8 & 16. &  0.0 &    3.7 &   -3656.9 &                  \\
     PV161 & 284.5 &  30.0 &  3998.5 &  5072.9 &  -24.8 &  8. &  0.0 &    4.9 &    -457.1 &                  \\
     PV162 & 307.7 &  30.0 &  6094.4 &  6769.7 &  -24.8 &  8. &  0.0 &    2.8 &    -457.1 &                  \\
\end{tabular}
\end{table*}
\medskip

\section*{
Acknowledgements} 
This work was supported by PPARC
(Grant no. GR/K97828).

\end{document}